\documentclass[11pt]{article}%
\usepackage{amsmath}
\usepackage{amsfonts}
\usepackage{amssymb}
\usepackage{amsthm}
\usepackage{graphicx}
\usepackage[shortlabels]{enumitem}%
\providecommand{\U}[1]{\protect\rule{.1in}{.1in}}
\providecommand{\U}[1]{\protect\rule{.1in}{.1in}}
\begin{document}

\begin{center}

{\Large \textbf{Generating M-indeterminate probability densities by way of
quantum mechanics}}

\bigskip

\textbf{Rafael Sala Mayato$^{a}$, Patrick Loughlin$^{b}$, Leon Cohen$^{c*}$}

\bigskip$^{a}$ Departamento de F\'{\i}sica and IUdEA, Universidad de La
Laguna, La Laguna 38203, Tenerife, Spain\newline$^{b}$ Departments of
Bioengineering, and Electrical \& Computer Engineering, University of
Pittsburgh, Pittsburgh, PA 15261, USA\newline$^{c}$ Department of Physics,
Hunter College of the City University of New York, 695 Park Ave., New York, NY
10065, USA. *Corresponding author; email: leon.cohen@hunter.cuny.edu
\end{center}

\bigskip

\noindent\textbf{{Abstract:}} Probability densities that are not uniquely
determined by their moments are said to be \textquotedblleft
moment-indeterminate,\textquotedblright\ or \textquotedblleft
M-indeterminate." Determining whether or not a density is M-indeterminate, or
how to generate an M-indeterminate density, is a challenging problem with a
long history. Quantum mechanics is inherently probabilistic, yet the way in
which probability densities are obtained is dramatically different in
comparison to standard probability theory, involving complex wave functions
and operators, among other aspects. Nevertheless, the end results are standard
probabilistic quantities, such as expectation values, moments and probability
density functions. We show that the quantum mechanics procedure to obtain
densities leads to a simple method to generate an infinite number of
M-indeterminate densities. Different self-adjoint operators can lead to new
classes of M-indeterminate densities. Depending on the operator, the method
can produce densities that are of the Stieltjes class or new formulations that
are not of the Stieltjes class. As such, the method complements and extends
existing approaches and opens up new avenues for further development. The
method applies to continuous and discrete probability densities. A number of
examples are given.

\bigskip


\bigskip

\section{Introduction}


Let $X$ be a continuous real random variable with associated probability
density $P(x)$ ($x \in \mathbb{R}^1$) and moments $\text{E\negthinspace}\left[  X^{n}\right]  $,
where $n$ is a positive integer and $\text{E\negthinspace}\left[  .\right]  $
denotes the expected value of the argument. Many of the common densities in
statistics are uniquely determined by their moments; however, some are not.
Probability densities that are not uniquely determined by their moments are
said to be \textquotedblleft moment-indeterminate,\textquotedblright\ or
\textquotedblleft M-indeterminate.\textquotedblright\ The idea first arose
with an example by Stieltjes but the best known one is the log-normal density
\cite{feller,stoy4},%

\begin{equation}
{P}_{\text{LN}}{(x)=}\frac{1}{x\sqrt{2\pi}}{\exp}\left[  -\frac{({\ln}%
\ x)^{2}}{2}\right]  \qquad0<x<\infty\nonumber
\end{equation}
which has moments
\begin{equation}
\text{E\negthinspace}\left[  X^{n}\right]  =\int_{0}^{\infty}x^{n}%
{P}_{\text{LN}}{(x)dx}=e^{n^{2}/2}\nonumber
\end{equation}
In a landmark paper, Heyde \cite{heyde} explicitly showed that the densities
given by
\begin{equation}
P(x)={P}_{\text{LN}}{(x)}\left[  {1+\varepsilon\sin(k 2\pi\ln x)}\right]
,\qquad-1\leq{\varepsilon}\leq1 \label{m-ind-lognormal}%
\end{equation}
where $k \in\mathbb{I^{+}}$, have the same moments as $P_{\text{LN}}(x){.}$
Thus, the log-normal is M-indeterminate. Determining whether or not a given
density is M-indeterminate, or how to generate an M-indeterminate density, is
a challenging problem with a long history in probability theory
\cite{stoy4}-\cite{schmudgen}.

The \textquotedblleft Stieltjes class\textquotedblright\ of M-indeterminate
densities can be formulated as a parameterized family of functions
\cite{stoy1,stoy2}
\begin{equation}
P_{\epsilon}(x)=P(x)\left[  1+{\varepsilon}\,h(x)\right]  ,\quad
-1\leq{\varepsilon}\leq1 \label{Stieltjes}%
\end{equation}
where $P(x)$ is an M-indeterminate density, and $h(x)\ne0$
is a continuous, bounded ($|h(x)|\leq1$) function called a \textquotedblleft
perturbation function,\textquotedblright\ subject to the condition that the
product $h(x)P(x)$ has vanishing \textquotedblleft moments,\textquotedblright%
\[
\int_{-\infty}^{\infty}x^{n}h(x)P(x)\,dx=0,\quad n=0,1,2,...
\]
Stoyanov and Tolmatz have developed a powerful method for constructing
Stieltjes classes of M-indeterminate densities \cite{stoy3}. Briefly, the
method consists of specifying the scale $\delta$ and shift $s$ parameters for
some continuous, bounded function $g(x)$ with vanishing moments, such that
\[
h(x)=c\,\frac{g(\delta x-s)}{P(x)}%
\]
exists for all $x$, where $c$ is a normalization constant. There are several
candidate functions $g(x)$, and convex combinations of such functions can also
be used \cite{stoy3}. A number of examples of M-indeterminate densities, and
detailed discussion of the moment problem, can be found in the book by
Stoyanov \cite{stoy4}.

We provide a different perspective on the problem of generating
M-indeterminate densities that derives from the way in which probability
densities are obtained in quantum mechanics (QM). While QM is inherently
probabilistic, the mathematics underlying it are quite different in comparison
to standard probability theory, involving complex wave functions, operators,
transformation theory and other aspects. We show how one can generate
different classes of M-indeterminate densities by choosing different
self-adjoint operators.

Operators are central in QM and correspond to measurable quantities or
\textquotedblleft observables,\textquotedblright\ such as energy, position,
momentum, etc. Operator methods have also been used in standard probability
theory, mostly involving the differentiation operator, which arises in the
study of cumulants \cite{s,hald} and the Gram-Charlier series \cite{gram}. For
a comparison of the quantum and standard approach to obtaining probabilities,
see \cite{are,rules}.

Despite the unorthodox methods of QM, we emphasize that the densities obtained
are proper in the classical sense. Moreover, the mathematics of QM bring forth
new ideas and approaches that can inform standard probability theory. The aim
of this paper is to show that the procedure by which densities are obtained in
QM leads to a simple procedure for generating new classes of M-indeterminate
densities, that complements existing approaches and opens up new avenues for
further development.

\section{Preliminaries: Notation, densities and expectation values in QM}

To set forth our notation and because of its distinction from standard
probability theory, we briefly review how probability densities are obtained
in QM \cite{bohm,merz}.

Measureable physical quantities, such as the position or momentum of a
particle, are inherently probabilistic and associated with self-adjoint
operators in QM. The measurable quantities are obtained from the quantum
mechanical wave function, often denoted by $\psi(x)$, which is complex and
represents the full description of the physical system.

\bigskip

\textbf{\noindent Definition 1 } {\it For a continuous random variable} $X$ {\it and wave
function} $\psi(x)$ $(x\in \mathbb{R}^{1})$ {\it in} $L^{2}$, {\it the QM probability density
associated with} $\psi(x)$  {\it is}
\[
P(x)=\left\vert \psi(x)\right\vert ^{2}%
\]
{\it with normalization}
\begin{equation}
\int_{-\infty}^{\infty}P(x)\,dx=\int_{-\infty}^{\infty}\left\vert
\psi(x)\right\vert ^{2}\,dx=1\nonumber
\end{equation}

\bigskip

The unique aspect of QM is how densities of other variables are obtained,
which is done by solving an eigenvalue problem,
\begin{equation}
\mathbf{A}\,u(r,x)=r\,u(r,x) \label{eigvalprob}%
\end{equation}
where $\mathbf{A}$ is a self-adjoint operator that represents a physical
quantity, and the $r$'s are the eigenvalues, which are the random variables in
QM. The $u$'s are the corresponding eigenfunctions, which are complete and
orthogonal.
Accordingly, $\psi(x)$ as defined may be expanded as \cite{mors}
\begin{equation}
\psi(x)=\int_{-\infty}^{\infty}F(r)u(r,x)\,dr\nonumber
\end{equation}
where
\begin{equation}
F(r)=\int_{-\infty}^{\infty}\psi(x)u^{\ast}(r,x)\,dx \label{r-transform}%
\end{equation}
The function $F(r)$ is the representation of the function in the $r$-domain,
and we shall call it the \textit{r-transform} of $\psi(x)$. In writing Eq.
\eqref{eigvalprob}, we have assumed the variables are continuous; the discrete
case is considered in sec. \ref{discrete-case}. \label{note1}

\bigskip

\noindent\textbf{Definition 2} \textit{For a continuous random variable $R$ and
transform $F(r)$ $(r\in \mathbb{R}^{1})$ in $L^{2}$ as defined in Eq. \eqref{r-transform}, 
the QM probability density
associated with $F(r)$ is }
\begin{equation}
P(r)=\left\vert F(r)\right\vert ^{2}=\left\vert \int_{-\infty}^{\infty}%
\psi(x)u^{\ast}(r,x)\,dx\ \right\vert ^{2}%
\end{equation}

\bigskip

\noindent Note that $|F(r)|^{2}\,$ is a proper density, in that it is non-negative, and
properly normalized since $|\psi(x)|^{2}$ was normalized,
\begin{equation}
\int_{-\infty}^{\infty}|F(r)|^{2}\,dr=\int_{-\infty}^{\infty}|\psi
(x)|^{2}dx\ =1\ \nonumber
\end{equation}

\bigskip

\noindent\textbf{Lemma 1} \textit{Assume $R$ is a continuous random variable
with density} $|F(r)|^{2}$ ($r \in \mathbb{R}^1$) \textit{ and $F(r)$ given by Eq. \eqref{r-transform}. Then
the average value of $R$, defined by}
\begin{equation}
\text{E\negthinspace}\left[  R\right]  =\int_{-\infty}^{\infty}r|F(r)|^{2}%
\,dr\nonumber
\end{equation}
\textit{can be equivalently calculated in terms of $\psi(x)$ by way of}
\begin{equation}
\text{E\negthinspace}\left[  R\right]  =\int_{-\infty}^{\infty}\psi^{\ast
}(x)\mathbf{A}\psi(x)\,dx
\end{equation}
\textit{More generally, for a real function $g(R),$ its average value is given
by}
\begin{equation}
\text{E\negthinspace}\left[  g(R)\right]  =\int_{-\infty}^{\infty
}g(r)|F(r)|^{2}dr=\int_{-\infty}^{\infty}\psi^{\ast}(x)g(\mathbf{A}{)}%
\psi(x)\,dx \label{te168}%
\end{equation}

\begin{proof} See \cite{bohm,merz} and references therein.
\end{proof}

\section{Generating M-indeterminate densities}

We now make use of the procedures by which densities and expectation values
are obtained in QM to obtain new classes of M-indeterminate probability
densities. In what follows, $R$ represents a continuous real random variable
with associated probability density $P(r)$, and $\psi(x)$ is any complex
function satisfying the conditions of Theorem 1.

\bigskip

\noindent\textbf{Theorem 1} \textit{Let $\mathbf{A}$ be any self-adjoint
operator with corresponding eigenfunctions $u(r,x)$, for which the support of
$\mathbf{A}^{n}\psi(x)$ equals the support of $\psi(x)$, $\,n\in\mathbb{I^{+}%
}$. Further, let $\psi_{1}(x)\neq0$ and $\psi_{2}(x)\neq0$ be any normalized
generally complex functions in $L^{2}$ having disjoint supports $(\psi
_{1}(x)\psi_{2}(x)=0)$. Then, for the family of functions
\begin{equation}
\psi(x)=\frac{1}{\sqrt{2}}\left(  \psi_{1}(x)+e^{i\beta}\psi_{2}(x)\right)
,\quad-\pi\leq\beta\leq\pi\label{seedfun}%
\end{equation}
the classes of densities given by
\begin{equation}
P(r)=\left\vert \int_{-\infty}^{\infty}\psi(x)u^{\ast}(r,x)\,dx\ \right\vert
^{2}\label{r-densities}%
\end{equation}
depend on the parameter $\beta$ but the moments} $\text{E\negthinspace}\left[
R^{n}\right]  $ \textit{are independent of $\beta$. Hence, the classes of
densities $P(r)$ are M-indeterminate.}

\bigskip

\begin{proof}
Substituting $\psi(x)$ from Eq. \eqref{seedfun} into Eq.
\eqref{r-densities}, we have
\begin{align}
P(r)  &  =\left\vert \,\frac{1}{\sqrt{2}}\left(  F_{1}(r)+e^{i\beta}%
F_{2}(r)\right)  \right\vert ^{2}\,
\,  =\,\frac{1}{2}\left(  \left\vert \,F_{1}(r)\,\right\vert ^{2}%
\,+\,\left\vert \,F_{2}(r)\,\right\vert ^{2}\,+\,2\operatorname{Re}%
\{e^{i\beta}F_{1}^{\ast}(r)F_{2}(r)\}\right) \label{theorem}
\end{align}
where
\begin{equation}
F_{k}(r)=\int_{-\infty}^{\infty}\psi_{k}(x)\,\,u^{\ast}(r,x)\,dx,\ \ k=1,2 \nonumber
\end{equation}
Clearly, the densities $P(r)$ depend on $\beta$.
For the
moments, we have
\begin{equation}
\text{E\negthinspace}\left[  R^{n}\right]  =\int_{-\infty}^{\infty} r^{n}P(r)\,dr\,=\,\frac{1}%
{2}\int_{-\infty}^{\infty} r^{n}\left(  \left\vert \,F_{1}(r)\,\right\vert ^{2}\,+\,\left\vert
\,F_{2}(r)\,\right\vert ^{2}\,+\,2\operatorname{Re}\{e^{i\beta}F_{1}^{\ast
}(r)F_{2}(r)\}\right)  \,dr \label{moments-in-r}
\end{equation}
or equivalently, by Eq. \eqref{te168},
\begin{align}
\text{E\negthinspace}\left[  R^{n}\right] & =\int_{-\infty}^{\infty}\psi^{\ast}(x)\mathbf{A}%
^{n}\psi(x)dx \nonumber \\
& = \frac{1}{2}\int_{-\infty}^{\infty} \psi_1^{\ast}(x)\mathbf{A}^{n}\psi_1(x)
+ \psi_2^{\ast}(x)\mathbf{A}^{n}\psi_2(x) +  e^{i\beta}\,\psi_1^{\ast}(x)\mathbf{A}^{n}\psi_2(x)
+ e^{-i\beta}\, \psi_2^{\ast}(x)\mathbf{A}^{n}\psi_1(x) \,dx
\end{align}
Hence, in order for the moments
E$[R^{n}]$ to be independent of $\beta$, we must have
\begin{equation}
\int_{-\infty}^{\infty} e^{i\beta}\,\psi_1^{\ast}(x)\mathbf{A}^{n}\psi_2(x)
+ e^{-i\beta}\, \psi_2^{\ast}(x)\mathbf{A}^{n}\psi_1(x) \,dx =0 \label{m-indet-condition}
\end{equation}
This condition follows immediately by the fact that $\psi_1(x)$ and $\psi_2(x)$ have disjoint support, and the support of $\mathbf{A}^{n}\psi_k(x)$ equals the support of $\psi_k(x)$; therefore the integrands are zero.
\end{proof}

\bigskip

\noindent\textit{Remark 1} This theorem generates various classes of
M-indeterminate densities in three ways:

\begin{enumerate}
[(a)]

\item First, for a given $\psi_{1}(x)$, $\psi_{2}(x)$ and self adjoint
operator $\mathbf{A}$, one has a particular family of densities, with
different members defined by the value of the parameter $\beta$. This is
analogous to the lognormal family as parameterized by $\varepsilon$ in Eq.
\eqref{m-ind-lognormal}, or the Stieltjes class of Eq. \eqref{Stieltjes} for a
specific $P(x)$.

\item Second and more significantly, for a given $\psi_{1}(x)$ and $\psi
_{2}(x)$, different self-adjoint operators yield different classes of
M-indeterminate densities.
Examples of different classes generated by choosing different self-adjoint
operators are provided in the next section.

\item Analogously, for a given self-adjoint operator, different classes are
defined by choosing different non-overlapping functions $\psi_{1}(x)$ and
$\psi_{2}(x)$.

One way to obtain such functions is to use \textquotedblleft bump
functions,\textquotedblright\ which are compactly supported differentiable
functions \cite{Tu}; there is a variety of such functions, one example being
the family of functions
\[
f(x)=\left\{
\begin{array}
[c]{ll}%
e^{-\frac{1}{1-x^{2k}}}\, & |x|<1\\
0,\, & \mathrm{otherwise}%
\end{array}
\right.
\]
where $k$ is a positive integer. Hence, one can take $\psi_{1}(x)$ to be a
unit-normalized bump function, e.g., $\psi_{1}(x)=cf(x)$ (where $c$ is a
normalization constant), which equals zero for $|x|>1$. Then take $\psi
_{2}(x)$ to be a shifted and possibly also scaled version of $\psi_{1}$,
namely $\psi_{2}(x)=\sqrt{\alpha}\,\psi_{1}(\alpha(x\pm D))$ with $\alpha>0$
and $D>1+\frac{1}{\alpha}$, by which it follows that $\psi_{1}(x)\psi
_{2}(x)=0$, as required.
\end{enumerate}

\noindent\textit{Remark 2 } The specific wave function given by Eq.
\eqref{seedfun} was considered in the QM context to show that the moments of
position and momentum are independent of $\beta$
\cite{ahar1,ahar-book,semon-taylor}. This result is known as one of Aharonov's
paradoxes. The issue has been recently revisited in terms of the
characteristic function approach to generating densities \cite{sala}. For our
purposes here, $\psi(x)$ is any function satisfying the conditions of Theorem 1.

\section{Examples of Different Classes of Densities}

We now consider various self-adjoint operators and the corresponding
M-indeterminate densities. As noted in Remark 1, each operator generates a new class.

As prescribed, we start with the normalized function of Eq. \eqref{seedfun}.
For simplicity but with sufficient generality, let $\psi_{1}(x)$ be a bump
function with $\psi_{1}(x)=0$ for $x<0$ or $x>a>0$, and let $\psi_{2}%
(x)=\psi_{1}(x-D)$, with $D>a$ so that $\psi_{1}(x)\psi_{2}(x)=0$ as required.
Because of the generality in choosing $\psi_{1}(x)$, we obtain an unlimited
number of M-indeterminate densities $P(r)$ for each operator/class.

\subsection{Example 1}

Consider the class of densities generated by the operator
\begin{equation}
\mathbf{A}=\frac{1}{i}\frac{d}{dx}%
\end{equation}
We note that this operator is the momentum operator in the position
representation in quantum mechanics (with $\hbar=1$) \cite{bohm,merz}. As
previously mentioned, it also appears in the study of cumulants \cite{s,hald}
and the Gram-Charlier series \cite{gram}. A basic property of the operator is
that it shifts a function,
\begin{equation}
e^{ir\mathbf{A}}f(x)=f(x+r)\nonumber
\end{equation}

Solving the eigenvalue problem for $\mathbf{A}$ yields the eigenfunctions
\begin{equation}
u(r,x)\,=\frac{1}{\sqrt{2\pi}}\,\,e^{irx}\nonumber
\end{equation}
Therefore, we have
\begin{align}
F(r)  &  =\frac{1}{\sqrt{2\pi}}\int_{-\infty}^{\infty}\psi(x)\,\,e^{-irx}%
dx\,\nonumber\\
&  =\,\frac{1}{\sqrt{2}}\left(  F_{1}(r)+e^{i\beta}F_{2}(r)\right) \nonumber\\
&  =\frac{1}{\sqrt{2}}F_{1}(r)(1+e^{i(\beta-rD)})\nonumber
\end{align}
where
\[
F_{1}(r)=\frac{1}{\sqrt{2\pi}}\int_{0}^{a}\psi_{1}(x)\,\,e^{-irx}dx\,
\]
and
\begin{align}
F_{2}(r)  &  =\frac{1}{\sqrt{2\pi}}\int_{D}^{a+D}\psi_{2}(x)\,\,e^{-irx}%
dx\nonumber\\
&  =\frac{1}{\sqrt{2\pi}}\int_{0}^{a}\psi_{1}(x)\,\,e^{-ir(x+D)}dx\nonumber\\
&  =e^{-irD}F_{1}(r)\nonumber
\end{align}

Accordingly, the probability density associated with $F(r)$ is
\begin{equation}
P(r)\,=\,\left\vert F(r)\right\vert ^{2}\,=\,\left\vert F_{1}(r)\right\vert
^{2}\left[  1+\cos\left(  rD-\beta\right)  \right]  \label{eq11}%
\end{equation}
which, as anticipated from Theorem 1, depends on $\beta$. However, the moments
E$[R^{n}]$ are independent of $\beta$ and are given by
\begin{align}
\text{E}[R^{n}]  &  =\int_{-\infty}^{\infty}r^{n}\,\left\vert F_{1}%
(r)\right\vert ^{2}\left[  1+\cos\left(  rD-\beta\right)  \right]
\,dr\nonumber\\
&  =\int_{-\infty}^{\infty}\psi^{\ast}(x)\mathbf{A}^{n}\psi(x)\,dx\nonumber\\
&  =\frac{1}{2}\int_{-\infty}^{\infty}\left(  \psi_{1}^{\ast}(x)+e^{-i\beta
}\psi_{1}^{\ast}(x-D)\right)  \left(  \frac{1}{i}\frac{d}{dx}\right)
^{n}\left(  \psi_{1}(x)+e^{i\beta}\psi_{1}(x-D)\right)  \,dx\nonumber\\
&  =\frac{1}{2}\int_{-\infty}^{\infty}\psi_{1}^{\ast}(x)\left(  \frac{1}%
{i}\frac{d}{dx}\right)  ^{n}\psi_{1}(x)\,dx\,+\,\frac{1}{2}\int_{-\infty
}^{\infty}\psi_{1}^{\ast}(x-D)\left(  \frac{1}{i}\frac{d}{dx}\right)  ^{n}%
\psi_{1}(x-D)\,dx\,\nonumber\\
&  \quad+\operatorname{Re}\left(  e^{-i\beta}\int_{-\infty}^{\infty}\psi
_{1}^{\ast}(x-D)\left(  \frac{1}{i}\frac{d}{dx}\right)  ^{n}\psi
_{1}(x)\,dx\right) \nonumber
\end{align}
But, since $D>a>0$ and $\psi_{1}(x)=0$ for $x \notin(0,a)$, it follows that
\begin{equation}
\int_{-\infty}^{\infty}\psi_{1}^{\ast}(x-D)\left(  \frac{1}{i}\frac{d}%
{dx}\right)  ^{n}\psi_{1}(x)\ dx=0\nonumber
\end{equation}
Hence, the moments are
\begin{align}
\text{E\negthinspace}\left[  R^{n}\right]   &  =\frac{1}{2}\int_{-\infty
}^{\infty}\psi_{1}^{\ast}(x)\left(  \frac{1}{i}\frac{d}{dx}\right)  ^{n}%
\psi_{1}(x)\,dx\,+\,\frac{1}{2}\int_{-\infty}^{\infty}\psi_{1}^{\ast
}(x-D)\left(  \frac{1}{i}\frac{d}{dx}\right)  ^{n}\psi_{1}(x-D)\,dx\nonumber\\
&  =\int_{-\infty}^{\infty}\psi_{1}^{\ast}(x)\left(  \frac{1}{i}\frac{d}%
{dx}\right)  ^{n}\psi_{1}(x)\,dx\nonumber
\end{align}
which do not depend on $\beta$ and therefore all the densities $P(r)$
parameterized by $\beta$ per Eq. \eqref{eq11} are M-indeterminate, as has been
shown previously \cite{semon-taylor, sala}. Note that Eq. \eqref{eq11} is of
the form of a Stieltjes class, Eq. \eqref{Stieltjes}; however, as the
following example shows, other self-adjoint operators can lead to new classes
of M-indeterminate densities that can not be expressed in the form of a
Stieltjes class.

\subsection{Example 2}

As a second example of a class of M-indeterminate densities, for the operator
we take \cite{cohen-weyl}
\begin{equation}
\mathbf{A}=cx+\frac{1}{i}\frac{d}{dx}%
\end{equation}
where $c$ is a real number. This operator is the sum of the position plus
momentum operators, properly dimensionalized. A property of this operator is
that
\begin{equation}
e^{ir\mathbf{A}}f(x)=e^{icrx-icr^{2}/2}f(x-r)\nonumber
\end{equation}
The eigenfunctions are
\begin{equation}
u(r,x)={\frac{1}{\sqrt{2\pi}}}\,e^{i(rx-cx^{2}/2)}\nonumber
\end{equation}
Hence, the \textit{$r$-transform} here is given by \cite{cohen-weyl}
\[
F(r)=\frac{1}{\sqrt{2\pi}}\int_{-\infty}^{\infty}\psi(x)\,\,e^{-irx+icx^{2}%
/2}dx
\]
For $\psi(x)$ as defined by Eq. \eqref{seedfun}, we thus have
\[
F(r)=\frac{1}{\sqrt{2}}\left(  F_{1}(r)+e^{i\beta}F_{2}(r)\right)
\]
where
\begin{equation}
F_{1}(r)=\frac{1}{\sqrt{2\pi}}\int_{0}^{a}\,e^{-irx+icx^{2}/2}\,\psi
_{1}(x)dx\nonumber
\end{equation}
and
\begin{align}
F_{2}(r)  &  =\frac{1}{\sqrt{2\pi}}\int_{D}^{D+a}e^{-irx+icx^{2}/2}\psi
_{2}(x)dx\,=\frac{1}{\sqrt{2\pi}}\int_{D}^{D+a}e^{-irx+icx^{2}/2}\psi
_{1}(x-D)dx\nonumber\\
&  =\frac{1}{\sqrt{2\pi}}e^{-irD+icD^{2}/2}\int_{0}^{a}e^{-i(r-cD)x+icx^{2}%
/2}\psi_{1}(x)dx\nonumber\\
&  =e^{-irD+icD^{2}/2}F_{1}(r-cD)\nonumber
\end{align}
Hence,
\begin{equation}
F(r)=\,\frac{1}{\sqrt{2}}\left[  F_{1}(r)+e^{i\beta}e^{-irD+icD^{2}/2}%
F_{1}(r-cD)\right] \nonumber
\end{equation}
and the probability density associated with $F(r)$ is
\begin{align}
P(r)  &  =\left\vert F(r)\right\vert ^{2}\nonumber\\
&  =\,\frac{1}{2}\left[  \left\vert F_{1}(r)\right\vert ^{2}+\left\vert
F_{1}(r-cD)\right\vert ^{2}+2\operatorname{Re}\{F_{1}^{\ast}(r)e^{i\beta
}e^{-irD+icD^{2}/2}F_{1}(r-cD)\}\right]
\end{align}
Here again, while the densities $P(r)$ depend on $\beta$, the moments do not
by virtue of Eq. \eqref{m-indet-condition}, and hence these densities are
M-indeterminate. Note that this case reduces to the previous case, Eq.
\eqref{eq11}, for $c=0$. However, for $c\neq0$, we can not express these
M-indeterminate densities in the form of Eq. \eqref{Stieltjes}.

\subsection{Example 3}

The operator
\begin{equation}
\mathbf{A}\,=\,\frac{1}{2i}\left(  x\frac{d}{dx}+\frac{d}{dx}x\right)
=\,\frac{1}{i}\left(  x\frac{d}{dx}+\frac{1}{2}\right)  =\frac{1}{i}\left(
\frac{d}{dx}x-\frac{1}{2}\right)
\end{equation}
arises in the compression or dilation of functions. In particular,
\cite{scale}%
\begin{equation}
e^{ir\mathbf{A}\,}f(x)=e^{r/2}f(e^{r}x)\nonumber
\end{equation}
The compression/dilation factor is $e^{r}$; for $0<e^{r}<1$, the function
$f(x)$ is dilated, while for $e^{r}>1$, the function is compressed. The factor
$e^{r/2}$ preserves normalization of $f(x)$:
\begin{equation}
\int_{-\infty}^{\infty}\left\vert f(x)\right\vert ^{2}dx=\int_{-\infty
}^{\infty}\left\vert e^{r/2}f(e^{r}x)\right\vert ^{2}dx=1\nonumber
\end{equation}

Solving the eigenvalue problem for $\mathbf{A}$ yields the eigenfunctions
\begin{equation}
u(r,x)\,=\,\,\frac{1}{\sqrt{2\pi}}\,\frac{\,e^{ir\ln x}}{\sqrt{x}}\ ,\quad
x\,\geq0\nonumber
\end{equation}
Hence, for one sided $\psi(x)$ we have
\begin{equation}
F(r)=\int_{0}^{\infty}\,\psi(x)\,\,u^{\ast}(r,x)\,\,dx=\,\,\,\frac{1}%
{\sqrt{2\pi}}\int_{0}^{\infty}\frac{\,e^{-ir\ln x}}{\sqrt{x}}\,\psi
(x)dx\nonumber
\end{equation}
which we note is a Mellin transform with argument $-ir+1/2.$

Now for our problem,
\begin{equation}
F(r)=\,\frac{1}{\sqrt{2}}\left(  F_{1}(r)+e^{i\beta}F_{2}(r)\right) \nonumber
\end{equation}
where
\begin{equation}
F_{1}(r)=\frac{1}{\sqrt{2\pi}}\int_{0}^{a}\,\frac{\,e^{-ir\ln x}}{\sqrt{x}%
}\,\psi_{1}(x)dx\,\nonumber
\end{equation}
and
\begin{align}
F_{2}(r)  &  =\frac{1}{\sqrt{2\pi}}\int_{D}^{D+a}\psi_{1}(x-D)\frac
{\,e^{-ir\ln x}}{\sqrt{x}}\,dx\nonumber\\
&  =\frac{1}{\sqrt{2\pi}}\int_{0}^{a}\psi_{1}(x)\frac{\,e^{-ir\ln(x+D)}}%
{\sqrt{x+D}}\,dx\nonumber
\end{align}
Therefore%
\begin{align}
F(r)  &  =\frac{1}{\sqrt{2}}\left(  \frac{1}{\sqrt{2\pi}}\int_{0}^{a}%
\,\frac{\,e^{-ir\ln x}}{\sqrt{x}}\,\psi_{1}(x)dx\,+e^{i\beta}\frac{1}%
{\sqrt{2\pi}}\int_{0}^{a}\psi_{1}(x)\frac{\,e^{-ir\ln(x+D)}}{\sqrt{x+D}%
}\,dx\right) \nonumber\\
&  =\frac{1}{\sqrt{2}}\frac{1}{\sqrt{2\pi}}\int_{0}^{a}\,\left(
\frac{\,e^{-ir\ln x}}{\sqrt{x}}+e^{i\beta}\frac{\,e^{-ir\ln(x+D)}}{\sqrt{x+D}%
}\right)  \,\psi_{1}(x)dx\nonumber
\end{align}
The probability densities associated with $F(r)$ are given by%
\begin{equation}
P(r)=\left\vert F(r)\right\vert ^{2}\,=\,\frac{1}{4\pi}\left\vert \int_{0}%
^{a}\,\left(  \frac{\,e^{-ir\ln x}}{\sqrt{x}}+e^{i\beta}\frac{\,e^{-ir\ln
(x+D)}}{\sqrt{x+D}}\right)  \,\psi_{1}(x)dx\,\right\vert ^{2}%
\end{equation}
As before, the densities depend on $\beta$, however the moments do not which
is straightforward to show via the operator approach of Eq. \eqref{te168}.

It is of interest to contrast the operator approach with the calculation of
the moments in the usual way, namely
\begin{align}
\text{E\negthinspace}\left[  R^{n}\right]   &  =\int_{-\infty}^{\infty}%
r^{n}\left\vert F(r)\right\vert ^{2}dr\,=\, \frac{1}{4\pi}\int_{-\infty
}^{\infty}r^{n}\left\vert \int_{0}^{a}\,\left(  \frac{\,e^{-ir\ln x}}{\sqrt
{x}}+e^{i\beta}\frac{\,e^{-ir\ln(x+D)}}{\sqrt{x+D}}\right)  \,\psi
_{1}(x)dx\right\vert ^{2}dr\nonumber\\
&  =\frac{1}{4\pi}\int_{-\infty}^{\infty}r^{n}\int_{0}^{a}\,\int_{0}%
^{a}\left(  \frac{\,e^{-ir\ln x}}{\sqrt{x}}+e^{i\beta}\frac{\,e^{-ir\ln(x+D)}%
}{\sqrt{x+D}}\right)  \left(  \frac{\,e^{ir\ln x^{\prime}}}{\sqrt{x^{\prime}}%
}+e^{-i\beta}\frac{\,e^{ir\ln(x^{\prime}+D)}}{\sqrt{x^{\prime}+D}}\right)
\psi_{1}^{\ast}\,(x^{\prime})\psi_{1}(x)dxdx^{\prime}\,dr\nonumber
\end{align}
Upon expanding we obtain
\begin{align}
\text{E\negthinspace}\left[  R^{n}\right]   &  =\frac{1}{4\pi}\int_{-\infty
}^{\infty}r^{n}\left\vert \int_{0}^{a}\,\frac{\,e^{-ir\ln x}}{\sqrt{x}}%
\psi_{1}(x)dx\right\vert ^{2}\,dr+\frac{1}{4\pi}\int_{-\infty}^{\infty}%
r^{n}\int_{0}^{a}\,\left\vert \int_{0}^{a}\frac{\,e^{-ir\ln(x+D)}}{\sqrt{x+D}%
}\psi_{1}(x)dx\right\vert \,dr\nonumber\\
&  + \frac{1}{4\pi} \int_{-\infty}^{\infty}r^{n}\int_{0}^{a}\,\int_{0}%
^{a}\,\left(  e^{-i\beta} \frac{\,e^{-ir\ln x}}{\sqrt{x}}\frac{\,e^{ir\ln
(x^{\prime}+D)}}{\sqrt{x^{\prime}+D}}+e^{i\beta}\frac{\,e^{ir\ln x^{\prime}}%
}{\sqrt{x^{\prime}}}\frac{\,e^{-ir\ln(x+D)}}{\sqrt{x+D}}\right)  \psi
_{1}^{\ast}\,(x^{\prime})\psi_{1}(x)dxdx^{\prime}dr\nonumber
\end{align}
The first two terms are independent of $\beta$. Therefore, to show that the
moments are independent of $\beta$, we need to show that
\begin{equation}
\label{eg3-cross-term}\int_{-\infty}^{\infty}\int_{0}^{a}\,\int_{0}^{a}%
\,r^{n}\left(  e^{-i\beta}\frac{\,e^{-ir\ln x}}{\sqrt{x}}\frac{\,e^{ir\ln
(x^{\prime}+D)}}{\sqrt{x^{\prime}+D}}+e^{i\beta}\frac{\,e^{ir\ln x^{\prime}}%
}{\sqrt{x^{\prime}}}\frac{\,e^{-ir\ln(x+D)}}{\sqrt{x+D}}\right)  \psi
_{1}^{\ast}\,(x^{\prime})\psi_{1}(x)dxdx^{\prime}dr\,=\,0
\end{equation}
for $D>a>0$ and $\psi_{1}(x)=0,\,\, x \notin(0,a)$.

Consider
\begin{align}
I  &  =\int_{-\infty}^{\infty}\int_{0}^{a}\,\int_{0}^{a}\,r^{n}\left(
\frac{\,e^{-ir\ln x}}{\sqrt{x}}\frac{\,e^{ir\ln(x^{\prime}+D)}}{\sqrt
{x^{\prime}+D}}\right)  \psi_{1}^{\ast}\,(x^{\prime})\psi_{1}(x)dxdx^{\prime
}\,dr\nonumber\\
&  =\,\int_{-\infty}^{\infty}\int_{0}^{a}\,\int_{D}^{D+a}\,r^{n}%
\frac{\,e^{ir\ln(x^{\prime}/x)}}{\sqrt{xx^{\prime}}}\psi_{1}^{\ast
}\,(x^{\prime}-D)\psi_{1}(x)dx^{\prime}\,dx\,dr\nonumber
\end{align}
Making the variable substitution $y=\ln\left(  x^{\prime}/x\right)  $ with
$dx^{\prime}=xe^{y}dy$ we obtain
\begin{align}
I  &  =\int_{0}^{a}\,\int_{\ln(D/x)}^{\ln((D+a)/x)}\,\int_{-\infty}^{\infty
}r^{n}\frac{\,e^{iry}}{\sqrt{x^{2}e^{y}}}\psi_{1}^{\ast}(xe^{y}-D)\psi
_{1}(x)\,xe^{y}dr\,dy\,dx\nonumber\\
&  =\int_{0}^{a}\,\int_{\ln(D/x)}^{\ln((D+a)/x)}\,\int_{-\infty}^{\infty}%
r^{n}\,e^{iry}\,\psi_{1}^{\ast}(xe^{y}-D)\psi_{1}(x)\,e^{y/2}%
\,dr\,dy\,dx\nonumber\\
&  =2\pi\,\int_{0}^{a}\,\int_{\ln(D/x)}^{\ln((D+a)/x)}\,\left\{  \left(
\frac{1}{i}\frac{d}{dy}\right)  ^{n}\delta(y)\right\}  \psi_{1}^{\ast}%
(xe^{y}-D)\,e^{y/2}\,\psi_{1}(x)\,dy\,dx\nonumber\\
&  =2\pi\,\int_{0}^{a}\,i^{n}\left.  \left\{  \left(  \frac{d}{dy}\right)
^{n}\psi_{1}^{\ast}(xe^{y}-D)\,e^{y/2}\right\}  \right\vert _{_{y=0}}\psi
_{1}(x)\,dx\nonumber
\end{align}
Evaluation leads to
\begin{equation}
I=2\pi\int_{0}^{a}\,\sum_{k=0}^{n}\frac{i^{n}x^{k}}{2^{n-k}}\binom{n}{k}%
\,\psi_{1}^{\ast(k)}(x-D)\,\psi_{1}(x)\,dx=0\nonumber
\end{equation}
where the last step follows since $\psi_{1}^{\ast(k)}(x-D)\,\psi_{1}(x)=0$ for
$D>a$. By an identical derivation, we obtain
\begin{equation}
\int_{-\infty}^{\infty}\int_{0}^{a}\,\int_{0}^{a}\,r^{n}\left(  \frac
{\,e^{ir\ln x^{\prime}}}{\sqrt{x^{\prime}}}\frac{\,e^{-ir\ln(x+D)}}{\sqrt
{x+D}}\right)  \psi_{1}^{\ast}\,(x^{\prime})\psi_{1}(x)dxdx^{\prime
}\,dr\,=\,0\nonumber
\end{equation}
for $D>a$. Hence Eq. \eqref{eg3-cross-term} holds, and the moments are
independent of $\beta$, as previously determined via the operator approach.

\subsection{Example 4}

\label{example4}

For the operator we take%
\begin{equation}
\mathbf{A}=-\frac{d^{2}}{dx^{2}}-\frac{1}{c}x
\end{equation}
where $c$ is a constant. This case is motivated by what is called the constant
force Hamiltonian in quantum mechanics. It consists of the kinetic energy
operator, $-\frac{d^{2}}{dx^{2}}$ (with $\hbar=1$ and mass equal to $1/2$),
plus the potential energy term, $-\frac{1}{c}x,$ which corresponds to a
constant force in the positive $x$ direction with magnitude $1/c$.

The eigenvalue problem is%
\begin{equation}
\left(  -\frac{d^{2}}{dx^{2}}-\frac{1}{c}x\right)  u(x,r)=r\,u(x,r)\nonumber
\end{equation}
where $r$ are the eigenvalues. The eigenvalue problem can be solved in any
representation. To solve this eigenvalue problem, it is simpler to convert to
the dual Fourier domain by defining%
\[
v(p,r)=\frac{1}{\sqrt{2\pi}}\int_{-\infty}^{\infty}e^{-ipx}u(x,r)\,dx
\]
in which case the eigenvalue problem becomes
\begin{equation}
\left(  p^{2}-\frac{i}{c}\frac{d}{dp}\right)  v(p,r)=r\,v(p,r)\nonumber
\end{equation}

The eigenfunctions, $v(p,r),$ are
\begin{equation}
v(p,r)=\sqrt{\frac{c}{2\pi}}\,e^{ic\left(  rp-p^{3}/3\right)  }\nonumber
\end{equation}
and the corresponding $u(x,r)$ are then given by%
\begin{align*}
u(x,r)  &  =\frac{1}{\sqrt{2\pi}}\int_{-\infty}^{\infty}e^{-ipx}v(p,r)\,dp\\
&  =\sqrt{\frac{c}{2\pi}}\int_{-\infty}^{\infty}e^{-ipx}e^{ic\left(
rp-p^{3}/3\right)  }\,dp
\end{align*}
These eigenfunctions can be expressed in terms of Airy functions but that is
not necessary for our considerations.

As with $\psi(x)$, we can expand $\varphi(p)$ in terms of the eigenfunctions
via
\begin{equation}
\varphi(p)\,=\sqrt{\frac{c}{2\pi}}\int_{-\infty}^{\infty}F(r)e^{ic\left(
rp-p^{3}/3\right)  }dr\nonumber
\end{equation}
where
\begin{equation}
F(r)\,=\sqrt{\frac{c}{2\pi}}\int_{-\infty}^{\infty}\varphi(p)\,e^{-ic\left(
rp-p^{3}/3\right)  }dp\nonumber
\end{equation}
Now for our problem, let $\varphi(p)$ consist of two functions with disjoint
support,
\begin{equation}
\varphi(p)=\frac{1}{\sqrt{2}}\left(  \varphi_{1}(p)+e^{i\beta}\varphi
_{2}(p)\right) \nonumber
\end{equation}
where $\varphi_{1}(p)=0$ for $p<0$ or $p>a>0$, and $\varphi_{2}(p)=\varphi
_{1}(p-D)$, with $D>a$ so that $\varphi_{1}(p)\varphi_{2}(p)=0$. Then, it
follows that
\begin{equation}
F(r)=\,\frac{1}{\sqrt{2}}\left(  F_{1}(r)+e^{i\beta}F_{2}(r)\right) \nonumber
\end{equation}
where
\begin{align*}
F_{1}(r)\,  &  =\sqrt{\frac{c}{2\pi}}\int_{0}^{a}\varphi_{1}(p)\,e^{ic\left(
rp-p^{3}/3\right)  }dp\\
F_{2}(r)  &  =\sqrt{\frac{c}{2\pi}}\int_{D}^{D+a}\varphi_{2}(p)e^{ic\left(
rp-p^{3}/3\right)  }dp\\
&  =\sqrt{\frac{c}{2\pi}}\int_{0}^{a}\varphi_{2}(p+D)\,e^{ic\left(
r(p+D)-(p+D)^{3}/3\right)  }dp
\end{align*}
The densities associated with $F(r)$ are
\begin{equation}
P(r)=\left\vert F(r)\right\vert ^{2}=\frac{1}{2}\left(  \left\vert
F_{1}(r)\right\vert ^{2}+\left\vert F_{2}(r)\right\vert ^{2}%
+2\operatorname{Re}\{F_{1}^{\ast}(r)e^{i\beta}F_{2}(r)\}\right)  \label{ex4}%
\end{equation}
which clearly depend on $\beta$ (even though $\varphi_{1}(p)\varphi_{2}(p)=0$,
in general $F_{1}(r)F_{2}(r)\neq0$).

The moments E$\left[  R^{n}\right]  $ are given by
\begin{equation}
\text{E\negthinspace}\left[  R^{n}\right]  =\int_{-\infty}^{\infty}%
r^{n}P(r)dr\,\,=\,\,\frac{1}{2}\int_{-\infty}^{\infty}r^{n}\left(  \left\vert
F_{1}(r)\right\vert ^{2}+\left\vert F_{2}(r)\right\vert ^{2}%
+2\operatorname{Re}\{F_{1}^{\ast}(r)e^{i\beta}F_{2}(r)\}\right)  dr\nonumber
\end{equation}
The first two terms are clearly independent of $\beta$; however, so is the
last term, by virtue of the fact that
\begin{equation}
\int_{-\infty}^{\infty}F_{1}^{\ast}(r)\,r^{n}\,F_{2}(r)\,dr=\int_{-\infty
}^{\infty}\varphi_{1}^{\ast}(p)\,\mathbf{A}^{n}\,\varphi_{2}%
(p)\,dp\,\,=0\nonumber
\end{equation}
which follows from the fact that $\varphi_{1}(p)$ and $\varphi_{2}(p)$ are
non-overlapping and the support of $\mathbf{A}^{n}\,\varphi_{2}(p)$ is equal
to the support of $\varphi_{2}(p)$. Accordingly, the densities $P(r)$ (Eq.
\eqref{ex4}) are M-indeterminate since they depend on the parameter $\beta$
but the moments do not.

\section{Discrete Case}

\label{discrete-case} The generation of M-indeterminate densities that are
discrete follows readily from the previous considerations, which we succinctly
present here. In this section, $R$ is a real discrete random variable that
assumes possible values $r_{1},r_{2},r_{3}$ ..... with corresponding
probabilities $P(r_{n})$.

For the case of self-adjoint operators $\mathbf{A}$ that give discrete
eigenvalues, we write the eigenvalue problem as
\begin{equation}
\mathbf{A}u_{n}(x)=r_{n}u_{n}(x)\nonumber
\end{equation}
The eigenfunctions $u_{n}(x)$ are complete and orthogonal, hence, any
continuous function in $L^{2}$ can be expanded as
\begin{equation}
\psi(x)=\sum_{n}c_{n}u_{n}(x)\nonumber
\end{equation}
where the coefficients $c_{n}$ are given by
\begin{equation}
c_{n}=\int_{-\infty}^{\infty}\psi(x)u_{n}^{\ast}(x)dx\nonumber
\end{equation}
and the normalization is such that
\begin{equation}
\int_{-\infty}^{\infty}|\psi(x)|^{2}dx\ =\sum_{n}\left\vert c_{n}\right\vert
^{2}=1\nonumber
\end{equation}

The eigenvalues are the discrete random variables and their associated
probabilities are given by
\begin{equation}
P(r_{n})=\left\vert c_{n}\right\vert ^{2}=\left\vert \int_{-\infty}^{\infty
}\psi(x)u_{n}^{\ast}(x)dx\right\vert ^{2} \label{discrete-pdf}%
\end{equation}
with expected value
\begin{equation}
\text{E\negthinspace}\left[  R\right]  =\sum_{n}r_{n}\left\vert c_{n}%
\right\vert ^{2}\nonumber
\end{equation}
By Lemma 1, we have equivalently that
\begin{equation}
\text{E\negthinspace}\left[  R\right]  =\int_{-\infty}^{\infty}\psi^{\ast
}(x)\mathbf{A}\psi(x)\,dx
\end{equation}
and more generally,
\begin{equation}
\text{E\negthinspace}\left[  g(R)\right]  =\sum_{n}g(r_{n})\left\vert
c_{n}\right\vert ^{2}=\int_{-\infty}^{\infty}\psi^{\ast}(x)g(\mathbf{A}{)}%
\psi(x)\,dx
\end{equation}

\noindent\textbf{Theorem 2 } \textit{The discrete probability density functions given by
Eq. \eqref{discrete-pdf} with $\psi(x)$ as defined in Eq. \eqref{seedfun} are
M-indeterminate for self-adjoint operators for which the support of
$\mathbf{A}^{n} \psi(x)$ equals the support of $\psi(x)$.}

\begin{proof}  The proof mirrors that of Theorem 1.
Briefly, we have
\begin{align}
P(r_{n}) &  =\left\vert \int_{-\infty}^{\infty}\psi(x)u_{n}^{\ast}(x)dx\right\vert ^{2}=\frac
{1}{2}%
\int_{-\infty}^{\infty}\int_{-\infty}^{\infty}
\left(  \psi_{1}^{\ast}(x^{\prime})+e^{-i\beta}\psi_{2}^{\ast}(x^{\prime
})\right)  u_{n}(x^{\prime})\left(  \psi_{1}(x)+e^{i\beta}\psi_{2}(x)\right)
u_{n}^{\ast}(x)dxdx^{\prime}\nonumber\\
&  =\frac{1}{2}(P_{1}(r_{n})+P_{2}(r_{n}))+\frac{1}{2}\left(  e^{i\beta}%
c_{n}^{\ast(1)}c_{n}^{(2)}+e^{-i\beta}c_{n}^{\ast(2)}c_{n}^{(1)}\right) \label{discrete-P(r)}
\end{align}
where%
\begin{equation}
P_{1}(r_{n})=\left\vert c_{n}^{(1)}\right\vert ^{2}\hspace{0.2in}%
;\hspace{0.2in}P_{2}(r_{n})=\left\vert c_{n}^{(2)}\right\vert ^{2}\nonumber
\end{equation}
\begin{equation}
c_{n}^{(1)}=\int_{-\infty}^{\infty}\psi_{1}(x)u_{n}^{\ast}(x)dx\ ;\qquad c_{n}^{(2)}=\int_{-\infty}^{\infty}\psi
_{2}(x)u_{n}^{\ast}(x)dx\nonumber
\end{equation}
In general the product
$c_{n}^{\ast(1)}c_{n}^{(2)}$ is not identically zero, in which case
$P(r_n)$ will depend on $\beta$.
On the other hand, as in the continuous case, because
$\psi_{1}(x)$ and $\psi_{2}(x)$ have disjoint support, the
moments are independent of $\beta$,
\begin{align*}
\text{E\negthinspace}\left[  R^{n}\right]  =%
{\displaystyle\sum\limits_{n=0}^{\infty}}
r^{n}\left\vert c_{n}\right\vert ^{2} &  =%
{\displaystyle\sum\limits_{n=0}^{\infty}}
r^{n}\left\vert \frac{1}{\sqrt{2}}(c_{n}^{(1)}+e^{i\beta}c_{n}^{(2)}%
)\right\vert ^{2}\\
&  =\frac{1}{2}\int_{-\infty}^{\infty}\left(  \psi_{1}^{\ast}(x)+e^{-i\beta}\psi_{2}^{\ast
}(x)\right)  \mathbf{A}^{n}\left(  \psi_{1}(x)+e^{i\beta}\psi_{2}(x)\right)
dx\\
&  =\frac{1}{2}\text{E\negthinspace}\left[  R^{n}\right]  _{1}+\frac{1}%
{2}\text{E\negthinspace}\left[  R^{n}\right]  _{2}+e^{-i\beta}\int_{-\infty}^{\infty}\psi
_{2}^{\ast}(x)\mathbf{A}^{n}\psi_{1}(x)dx+e^{i\beta}\int_{-\infty}^{\infty}\psi_{1}^{\ast
}(x)\mathbf{A}^{n}\psi_{2}(x)dx
\end{align*}
where E$[R^{n}]_{1}$ and E$[R^{n}]_{2}$ are the expectation values
\begin{align}
\text{E\negthinspace}\left[  R^{n}\right]  _{1}  & =%
{\displaystyle\sum\limits_{n=0}^{\infty}}
r^{n}\left\vert c_{n}^{(1)}\right\vert ^{2}=\int_{-\infty}^{\infty}\psi_{1}^{\ast}(x)\mathbf{A}%
^{n}\psi_{1}(x)dx\nonumber \\
\text{E\negthinspace}\left[  R^{n}\right]  _{2}\,  & =\,%
{\displaystyle\sum\limits_{n=0}^{\infty}}
r^{n}\left\vert c_{n}^{(2)}\right\vert ^{2}=\int_{-\infty}^{\infty}\psi_{2}^{\ast}(x)\mathbf{A}%
^{n}\psi_{2}(x)dx \nonumber
\end{align}
But since $\psi_1(x)\psi_2(x)=0$, Eq. \eqref{m-indet-condition} holds, by which it follows that
the moments are independent of $\beta$ and therefore all the (different)  $P(r_n)$ given by
Eq. \eqref{discrete-P(r)} have the same moments.
%
\end{proof}

\section{Conclusion}

M-indeterminate densities are those that are not uniquely determined by their
moments. Constructing such densities and/or determining whether or not a
density is M-determinate has historically been a challenging problem, although
many such densities and tests have been discovered since the issue was first
considered by Stieltjes.

From a quantum perspective, the issue first arose with the consideration of
two non-overlapping wave functions in position space as given by Eq.
\eqref{seedfun}. Aharonov \textit{et al.} showed that the moments of both the
position density and of the momentum density are independent of the parameter
$\beta$ \cite{ahar1,ahar-book,semon-taylor}. This latter result renders the
momentum density M-indeterminate, since the density itself does depend on
$\beta$ \cite{semon-taylor}. The analogous case of non-overlapping momentum
wave functions has also been considered \cite{wang-scully}.

We have shown that the unique way in which probability densities are obtained
in quantum mechanics gives rise to a simple procedure for constructing classes
of M-indeterminate densities. Namely, we start with a complete and orthogonal
set of functions $u(r,x)$ that are solutions of the eigenvalue problem for a
self-adjoint operator $\mathbf{A}$ with eigenvalues $r$,
\begin{equation}
\mathbf{A}u(r,x)=ru(r,x)\nonumber
\end{equation}
Then for the generally complex, continous function $\psi(x)$ one forms the
transform
\begin{equation}
F(r)=\int_{-\infty}^{\infty}\psi(x)u^{\ast}(r,x)\,dx\nonumber
\end{equation}
Normalizing such that $\int_{-\infty}^{\infty}\left\vert \psi(x)\right\vert
^{2}\,dx=1,$ it follows that $P(r)=\left\vert F(r)\right\vert ^{2}$ is a
proper probability density.

Different classes of densities $P(r)$ are obtained by choosing different
operators and/or functions $\psi(x)$. The densities in each class will be
M-indeterminate when the following conditions hold:

\begin{enumerate}
\item \label{c1} The function $\psi(x)$ is of the form $\psi(x)=\frac{1}%
{\sqrt{2}}\left(  \psi_{1}(x)+e^{i\beta}\psi_{2}(x)\right)  $, where $\psi
_{1}(x)$ and $\psi_{2}(x)$ are each normalized to one and, crucially, have
disjoint support:
\begin{equation}
\psi_{1}(x)\psi_{2}(x)=0\nonumber
\end{equation}

\item \label{c2} It follows that the function in the $r$-domain is
$F(r)=\int_{-\infty}^{\infty}\psi(x)u^{\ast}(r,x)\,dx\,=\,\frac{1}{\sqrt{2}%
}\left(  F_{1}(r)+e^{i\beta}F_{2}(r)\right)  $. \textit{However}, while the
functions $\psi_{1},\,\psi_{2}$ do not overlap in the $x$-domain, they
\textit{do} overlap in the $r$-domain:
\begin{equation}
F_{1}(r)F_{2}(r)\neq0\nonumber
\end{equation}

\item \label{c3} For any nonnegative integer $n$, $\mathbf{A}^{n}\psi_{k}(x)$
has the same support in $x$ as does $\psi_{k}(x)$. Together with condition 1
above, it therefore follows that
\[
\psi_{1}^{\ast}(x)\mathbf{A}^{n}\psi_{2}(x) \,=\,\psi_{2}^{\ast}%
(x)\mathbf{A}^{n}\psi_{1}(x) = 0
\]
This will be the case for (finite order) differential operators, and many
other operators as shown in sec. 4.
\end{enumerate}

Condition \ref{c2} means that the densities $P(r)$ will depend on the
parameter, $\beta.$ However, Conditions \ref{c1} and \ref{c3} render the
moments E$[R^{n}]$ independent of $\beta$, which follows readily from the
operator-procedure for calculating moments, namely Eq. \eqref{te168}, which we
re-state specifically in terms of moments,
\begin{equation}
\text{E\negthinspace}\left[  R^{n}\right]  =\int_{-\infty}^{\infty}%
r^{n}|F(r)|^{2}\,dr\,=\,\int_{-\infty}^{\infty}\psi^{\ast}(x)\mathbf{A}%
^{n}\psi(x)\,dx\nonumber
\end{equation}
One can also readily show that
\begin{equation}
\int_{-\infty}^{\infty}F_{1}^{\ast}(r)r^{n}F_{2}(r)\,dr\,=\,\int_{-\infty
}^{\infty}\psi_{1}^{\ast}(x)\mathbf{A}^{n}\psi_{2}(x)\,dx\nonumber
\end{equation}
by which it becomes clear, via Condition \ref{c3}, that the moments are
independent of $\beta$. Hence, the densites $P(r)$ are M-indeterminate.

\bigskip

\section{Acknowledgments}

R. Sala Mayato acknowledges funding by the Spanish MINECO and FEDER, grant
FIS2017-82855-P (MINE CO/ FEDER, UE), and by M. Payne.

\newpage
\end{document}